\documentclass[lettersize,journal]{IEEEtran}
\usepackage{amsmath,amsfonts}
\usepackage{algorithmic}
\usepackage{algorithm}
\usepackage{array}
\usepackage[font=small,labelfont=bf]{caption}
\usepackage{textcomp}
\usepackage{stfloats}
\usepackage{url}
\usepackage{verbatim}
\usepackage{graphicx}
\usepackage{cite}
\usepackage{multirow}
\usepackage[colorlinks,citecolor=blue]{hyperref}

\usepackage{tikz}
\usetikzlibrary{arrows.meta}
\usetikzlibrary{calc,decorations.markings}
\PassOptionsToPackage{dvipsnames,svgnames}{xcolor}   
\usetikzlibrary{arrows,calc}
\tikzset{
  on each segment/.style={
    decorate,
    decoration={
      show path construction,
      moveto code={},
      lineto code={
        \path [#1]
        (\tikzinputsegmentfirst) -- (\tikzinputsegmentlast);
      },
      curveto code={
        \path [#1] (\tikzinputsegmentfirst)
        .. controls
        (\tikzinputsegmentsupporta) and (\tikzinputsegmentsupportb)
        ..
        (\tikzinputsegmentlast);
      },
      closepath code={
        \path [#1]
        (\tikzinputsegmentfirst) -- (\tikzinputsegmentlast);
      },
    },
  },
  mid arrow/.style={postaction={decorate,decoration={
        markings,
        mark=at position .5 with {\arrow[#1]{stealth}}
      }}},
  quart arrow/.style={postaction={decorate,decoration={
        markings,
        mark=at position .15 with {\arrow[#1]{stealth}},
        mark=at position .95 with {\arrow[#1]{stealth}}
      }}},
}

\DeclareRobustCommand*{\IEEEauthorrefmark}[1]{%
  \raisebox{0pt}[0pt][0pt]{\textsuperscript{\footnotesize\ensuremath{#1}}}}

\begin{document}

%+++++++++++++++++++++++++++++++++++++++++++
\title{\LARGE Benchtop magnetic shielding for benchmarking atomic magnetometers}
% \author{Peter~James~Hobson, Niall~Holmes, Prashant~Patel, Ben~Styles, James~Chalmers, Chris~Morley, Alister~Davis, Michael~Packer, Thomas~Smith, Sintija~Raudonyte, Darragh~Holmes, Robert~Harrison, David~Woolger, Dominic~Sims, Matthew~Brookes, Richard~Bowtell, and Mark~Fromhold}
\author{P.~J.~Hobson\IEEEauthorrefmark{1, +},
N.~Holmes\IEEEauthorrefmark{1,2},
P.~Patel\IEEEauthorrefmark{3},
B.~Styles\IEEEauthorrefmark{3},
J.~Chalmers\IEEEauthorrefmark{3},
C.~Morley\IEEEauthorrefmark{1},
A.~Davis\IEEEauthorrefmark{1},
M.~Packer\IEEEauthorrefmark{1},
T.~X.~Smith\IEEEauthorrefmark{1},
S.~Raudonyte\IEEEauthorrefmark{3},
D.~Holmes\IEEEauthorrefmark{3},
R.~Harrison\IEEEauthorrefmark{3},
D.~Woolger\IEEEauthorrefmark{3},
D.~Sims\IEEEauthorrefmark{3},
M.~J.~Brookes\IEEEauthorrefmark{1,2},
R.~W.~Bowtell\IEEEauthorrefmark{1,2}, and
T.~M.~Fromhold\IEEEauthorrefmark{1} \\
\IEEEauthorblockA{\IEEEauthorrefmark{1}School of Physics and Astronomy, University of Nottingham, NG7 2RD, United Kingdom} \\
\IEEEauthorblockA{\IEEEauthorrefmark{2}Sir Peter Mansfield Imaging Centre, University of Nottingham, NG7 2RD, United Kingdom} \\
\IEEEauthorblockA{\IEEEauthorrefmark{3}Magnetic Shields Ltd, Staplehurst, TN12 0DS, United Kingdom} \\
\IEEEauthorblockA{\IEEEauthorrefmark{+}\href{mailto:peter.hobson@nottingham.co.uk}{peter.hobson@nottingham.ac.uk}}}

\maketitle

\begin{abstract} % 150 words
Here, a benchtop hybrid magnetic shield containing four mumetal cylinders and nine internal flexible printed circuit boards is designed, constructed, tested, and operated. The shield is designed specifically as a test-bed for building and operating ultra-sensitive quantum magnetometers. The geometry and spacing of the mumetal cylinders are optimized to maximize shielding efficiency while maintaining Johnson noise $\mathbf{<15}$~fT/$\mathbf{\sqrt{\mathbf{Hz}}}$. Experimental measurements at the shield's center show passive shielding efficiency of $\mathbf{\left(1.0\pm0.1\right){\times}10^6}$ for a $\mathbf{0.2}$~Hz oscillating field applied along the shield's axis.  The nine flexible printed circuit boards generate three uniform fields, which all deviate from perfect uniformity by $\mathbf{\leq0.5}$\% along $\mathbf{50}$\% of the inner shield axis, and five linear field gradients and one second-order gradient, which all deviate by $\mathbf{\leq4}$\% from perfect linearity and curvature, respectively, over measured target regions. Together, the target field amplitudes are adjusted to minimize the remnant static field along $\mathbf{40}$\% of the inner shield axis, as mapped using an atomic magnetometer. In this region, the active null reduces the norm of the magnitudes of the three uniform fields and six gradients by factors of $\mathbf{19.5}$ and $\mathbf{19.8}$, respectively, thereby reducing the total static field from $\mathbf{1.68}$~nT to $\mathbf{0.23}$~nT.
\end{abstract}


\bstctlcite{IEEEexample:BSTcontrol}

\section{Introduction}
An exceptionally low and controlled magnetic field is required to reduce noise in fundamental physics experiments~\cite{8704901,10.1063/1.4894158} and to benchmark ultra-sensitive quantum magnetometers, including those based on NV-centers~\cite{doi.org/10.1002/qute.202000111} and atomic vapors~\cite{10.1063/5.0062791,doi:10.1063/1.5098088,doi:10.1063/1.5042033}. In particular, zero-field Optically Pumped Magnetometers (OPMs)~\cite{10.1038/nature01484, Shah_2013} have diverse applications from functional neuroimaging~\cite{10.1016/j.neuroimage.2022.119084,BOTO2022119027} to rapid diagnostics of electric batteries~\cite{app10217864, Hu10667}, but require low static (no time variation) fields to reduce projection errors~\cite{BORNA2022118818} and nonlinearities in sensor gain.

External magnetic fields may be attenuated by enclosing a region with passive shielding material. For low frequency shielding, high permeability materials, like \emph{mumetal}, are used to divert magnetic flux. However, high permeability materials magnetize under applied fields, thereby limiting the shielding effect. Although this is mitigated by degaussing~\cite{9784868,doi:10.1063/1.4922671,doi:10.1063/1.4949516}, some remnant magnetization usually remains. Coil systems inside passive shields are used to null offsets induced by magnetization and cancel leakage fields. These coils may be designed to account for the electromagnetic distortion induced by their coupling to passive shielding~\cite{PhysRevApplied.15.054004,PhysRevApplied.15.064006,DiscretePaper,9815315,9873890}.

This article presents the design, construction, testing, and operation of a magnetic shield comprizing nested mumetal cylinders with end caps and internal active coils. The nested cylinders are of a high permeability and their geometries and spacings are optimized to maximize their shielding effectiveness and minimize the weight of the shield, while ensuring that there is a large internal usable volume for experimentation.  Entry hole positions are selected to maximize access to this useable volume without significantly diminishing the effectiveness of the passive shielding. The passive shielding performance is experimentally validated before nine active coils are constructed, housed on nested flexible Printed Circuit Boards (flex-PCBs), within the inner mumetal shield. These coils are designed to null static offsets due to magnetization and residual external magnetic fields which pass through the passive shielding. The coupling of the active and passive components is included \emph{a priori} in the design process to enhance the nulling process. The flex-PCBs are characterized in situ to validate the design procedure. Finally,  the coil currents applied to the flex-PCBs are tuned to null the residual field along the inner shield's axis as measured using a zero-field OPM.

\section{Passive shielding}\label{sec.shielding}
\begin{figure}[ht!]
\centering
\includegraphics[width=\columnwidth]{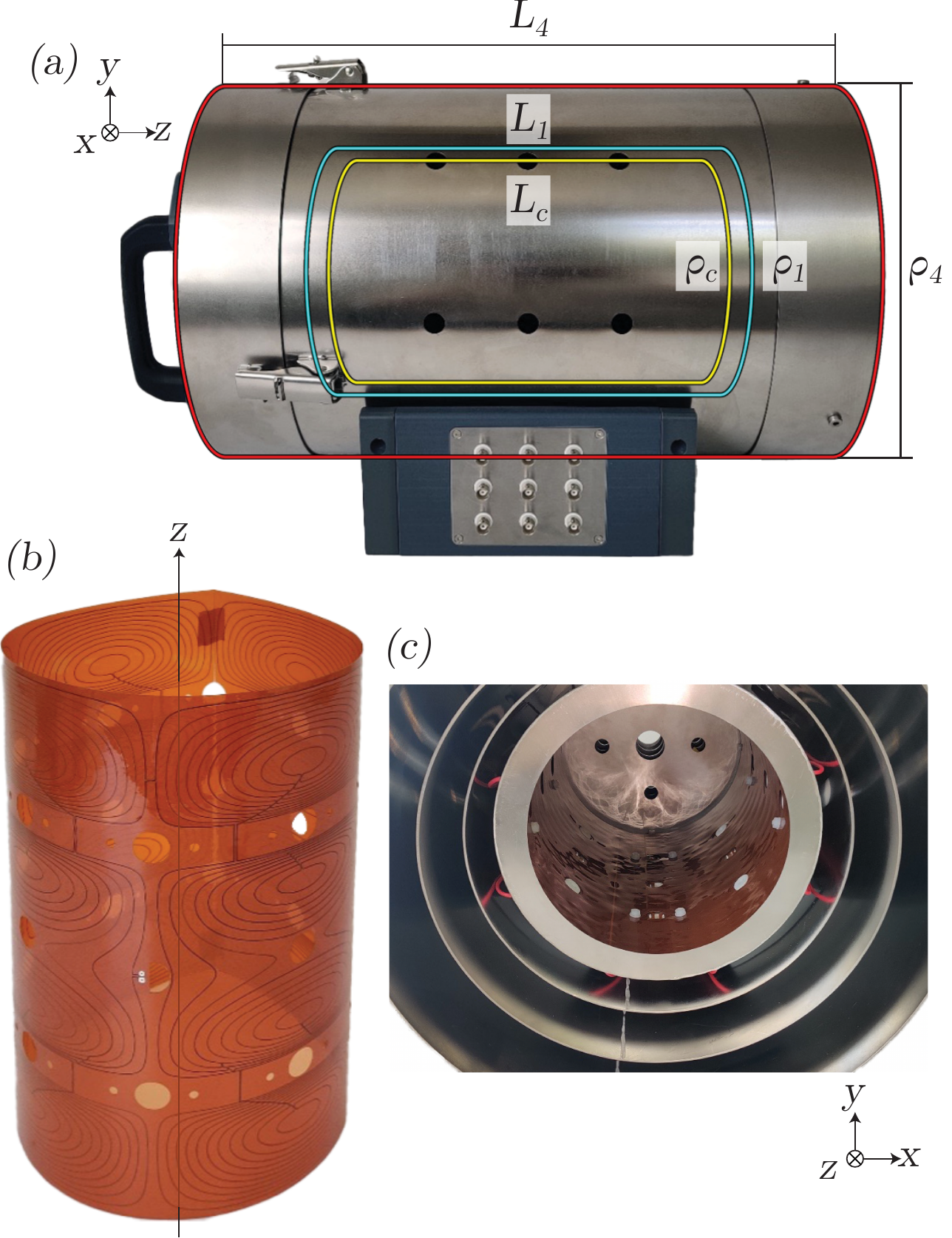}
\caption{The benchtop shield consists of four nested mumetal cylinders of outer radius $\rho_4=150$~mm and length $L_4=480$~mm [red] and inner radius $\rho_1=100$~mm and length $L_1=300$~mm [blue], which enclose a co-axial and co-centered set of rolled flex-PCBs of exterior radius $\rho_c=95$~mm and length $L_c=270$~mm [yellow]. (a) Side view of the shield and end caps, (b) a rolled PCB, and (c) multiple PCBs housed inside the shield with the end cap removed.}
\label{fig.shield}
\end{figure}
The passive shielding is constructed from four benchtop-sized nested co-axial and co-centered mumetal cylinders with access holes to allow optical access and cabling (Fig.~\ref{fig.shield}). Following Refs.~\cite{10.1117/12.2598780}~and~\cite{996017} the geometries of the exterior mumetal cylinders are optimized using the NGSA-II genetic algorithm~\cite{doi:10.1063/1.5131250}. This is detailed in appendix~\ref{app.gaopt}. The optimization returns the radii and lengths of the shield layers that maximize analytic approximations~\cite{Sumner_1987} for the shielding efficiency, $SE_{A,T}=\left|\mathbf{B}^{\mathrm{unshielded}}/\mathbf{B}^{\mathrm{shielded}}\right|$, along and transverse to the shield's axis ($A$/$T$) while minimizing the total volume of shielding material, $V$, which is proportional to the shield's weight. To ensure sufficient capacity for multiple atomic magnetometers, the inner cylinder dimensions are fixed, pre-optimization, to radius $\rho_1=100$~mm and length $L_1=300$~mm. The thickness of the inner mumetal layer is fixed to $0.5$~mm so that the shield-induced Johnson noise,  which is proportional to coil thickness~\cite{10.1063/1.2885711},  is reduced. The remaining shielding layers are fixed to $1.5$~mm thickness to balance shielding effectiveness with total weight. Access holes are selected manually to maximize optical access to the central half inner length and diameter of the inner shield cylinder, where the flex-PCB coils generate optimized magnetic fields, without blocking the wire patterns (see section \ref{sec.coils}). The center of each end cap has a $15$~mm radius access hole and is surrounded by four equally separated $7.5$~mm radius access holes which are separated by $50$~mm from the shield's axis. Additional holes of the same radius are placed in equally separated bands of eight at axial positions $z=[-65,0,65]$~mm in the cylindrical shield wall.

\begin{table}[ht!]
\caption{Axial and transverse shielding efficiency, $SE_A$ and $SE_T$, respectively, at the center of the benchtop shield subject to a spatially-uniform oscillatory field of frequency, $f$, and peak-to-peak amplitude $B_{0}=505$~$\mu\mathrm{T}$}.
    \centering
    \begin{tabular}{| c || c | c |}
    \hline
    $f$ (Hz) & $SE_A\times10^6$  & $SE_T\times10^6$ \\
    \hline
    \hline 
    $0.2$ & $1.0\pm0.1$ & $20.2\pm0.1$ \\
    \hline
    $1 $ & $1.1\pm0.1$ & $23.5\pm0.1$ \\
    \hline
    $10$ & $4.1\pm0.1$ & $39.2\pm0.1$ \\
    \hline
    $50$ & $4.5\pm0.1$ & $19.1\pm0.1$ \\
    \hline
    \end{tabular}
\label{tab.shieldstats}
\end{table}
\begin{figure}[htb!]
\begin{center}
   \begin{tabular}{c} %% tabular useful for creating an array of images 
         \includegraphics[width=\columnwidth]{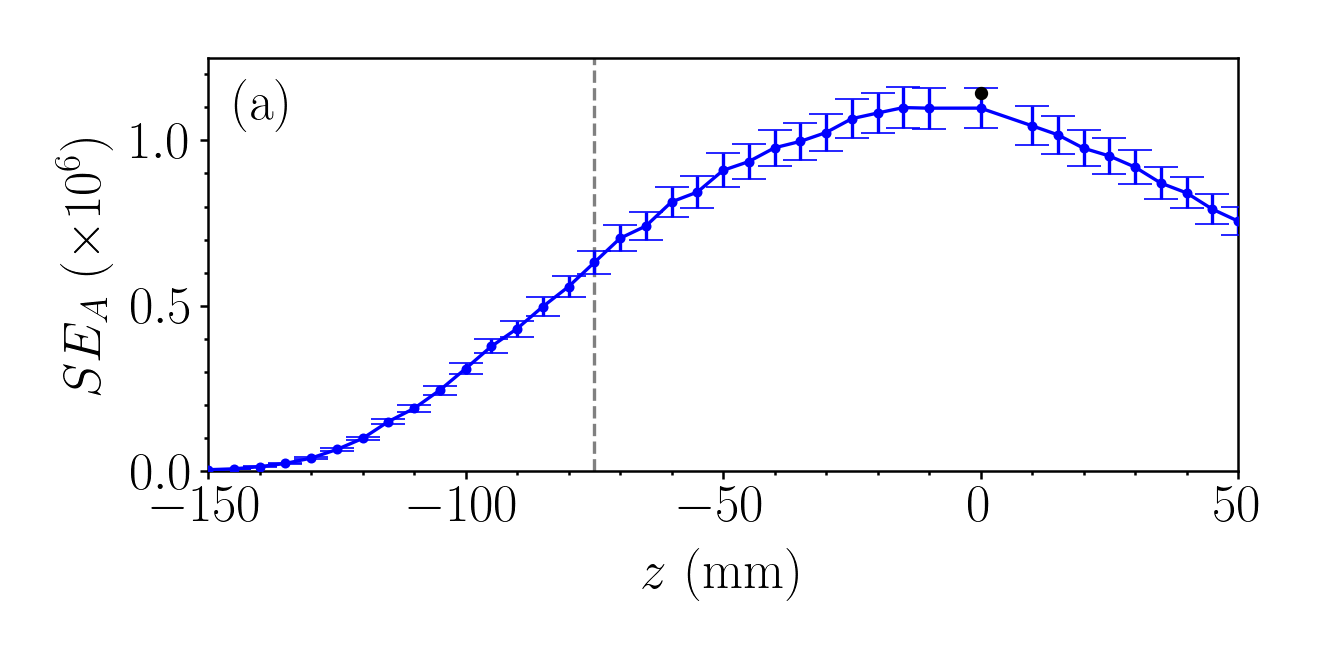} \\
         \vspace{-20pt}
         \includegraphics[width=\columnwidth]{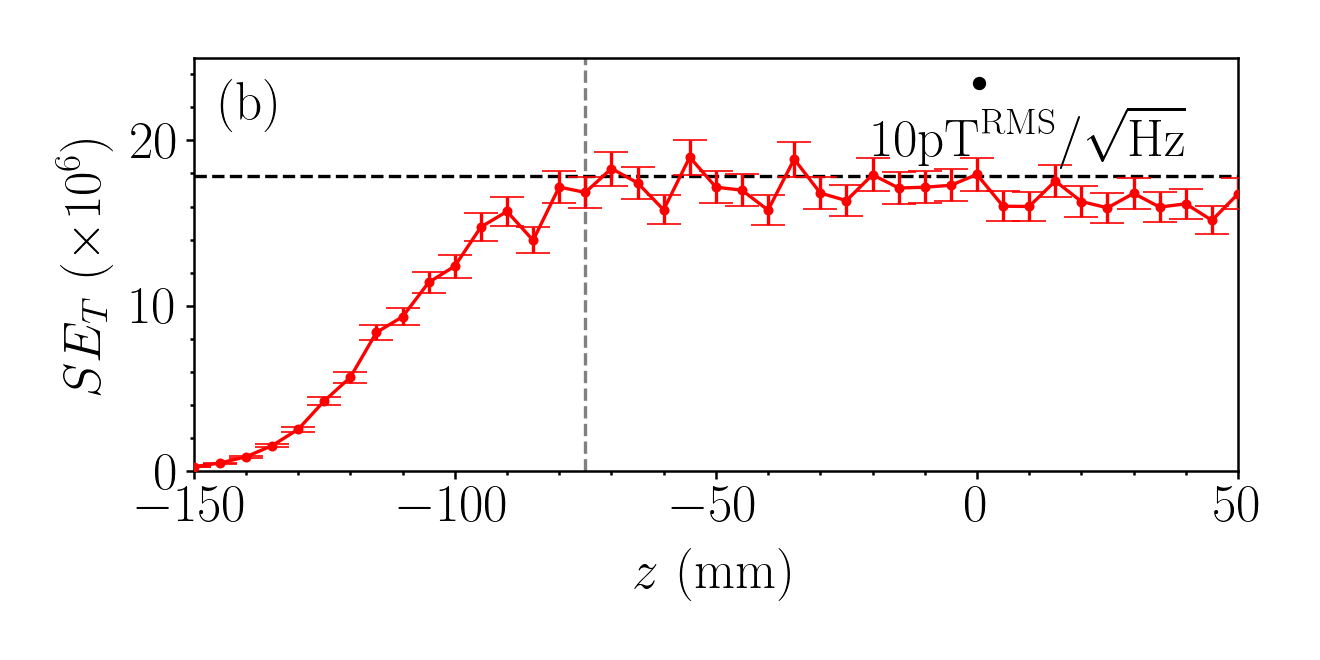}
    \end{tabular}
\end{center}
\vspace{5pt}
\caption{(a) Axial and (b) transverse shielding efficiency, $SE_A$ and $SE_T$, respectively, measured along the $z$-axis of the inner shield cylinder using a fluxgate magnetometer [red and blue] and an OPM [black] under a $f=1$~Hz sinusoidal drive field of peak-to-peak amplitude $B_{0}=505$~$\mu\mathrm{T}$ inside a MSR. Dashed grey lines show the edge of the target field region and dashed black lines [(b) only] show the noise limit of the fluxgate magnetometer, $10$~pT\textsuperscript{RMS}$\sqrt{\mathrm{Hz}}$.}
\label{fig.SE}
\end{figure}
\begin{figure}[htb!]
\begin{center}
\includegraphics[width=\columnwidth]{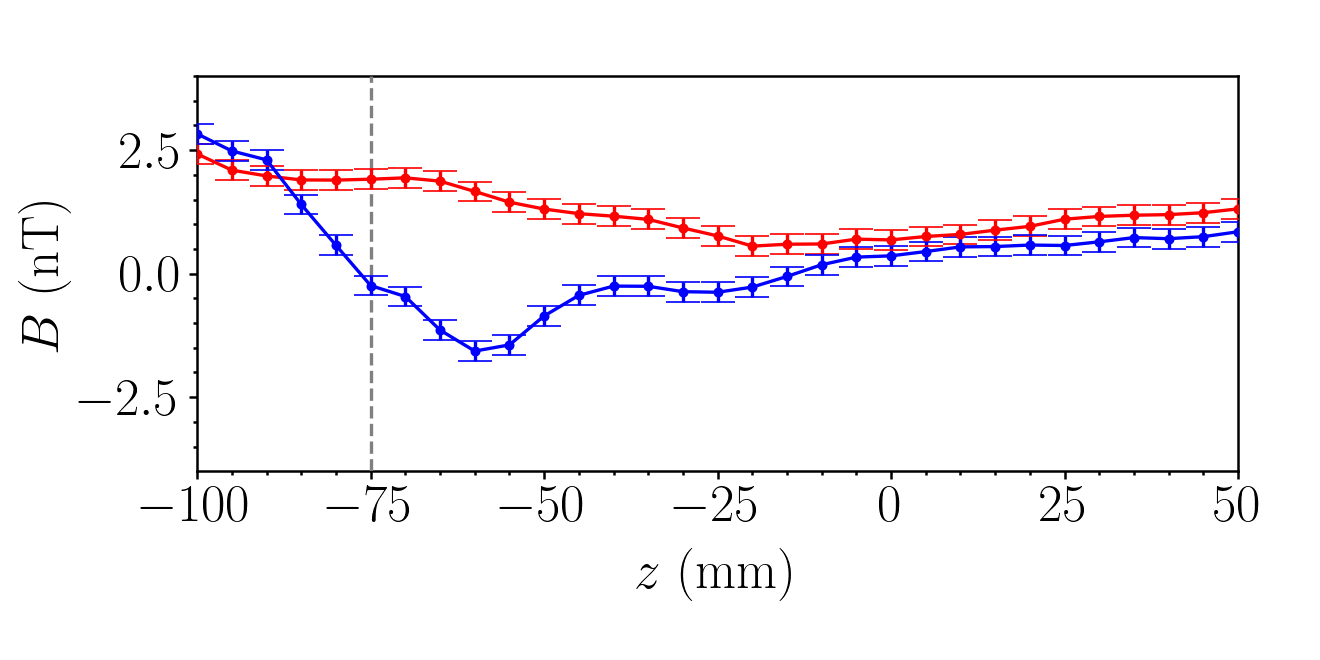}
\end{center}
\vspace{5pt}
\caption{Residual $\mathrm{norm}(B_x,B_y)$ and $B_z$ measured along the $z$-axis of the inner shield cylinder using a fluxgate magnetometer [red and blue] inside a MSR. Dashed grey line shows the edge of the target field region.}
\label{fig.mag}
\end{figure}
Next, we experimentally measure the passive shielding effectiveness of the shield by driving large uniform fields through it and measuring their attenuation inside the shield. The shield is degaussed whenever it is opened. During degaussing, a sinusoidal current with a peak-to-peak amplitude of $2$~A and at a frequency of $10$~Hz is driven for $10$~s through four loops that are wrapped along the inner shield's axis. The amplitude of the sinusoidal current is then ramped down linearly over $60$~s.  The drive current is generated using a National Instruments (NI) USB-6212 Data Acquisition module (DAQ) and is amplified using an AE Techron 7224 Power Amplifier, which is connected to a transformer to remove static offsets. The degaussing control parameters are determined by monitoring when the waveform of the induced voltage in a separate loop wrapped around the inner layer begins to distort during the fixed-amplitude phase. This provides the smallest energetic increments to remove residual magnetization during the ramping phase~\cite{10.1063/1.2713433}.

Sinusoidal fields of amplitude $252.5$~$\mu\mathrm{T}$ are applied to the shield by driving sinusoidal current of amplitude $5$~A through a $650$~mm radius Ferronato\textsuperscript{TM} circular Helmholtz cage (Serviciencia S.L.). The shield and Helmholtz cage are placed inside of a Magnetically Shielded Room (MSR) to minimize background fields so that the applied field is generated predominantly by the Helmholtz cage, maximizing the Signal-to-Noise Ratio (SNR). The Helmholtz cage generates independent uniform fields directed along each Cartesian coordinate axis that spatially deviate by $<1$\% from perfect uniformity in free space within a spherical volume that encompasses the entire inner shield. The same electronics as the degaussing system are used to drive the coils. To determine $SE_{A,T}$ at different drive frequencies, we calculate the Fast Fourier Transform (FFT) of the field at the center of the shield. The field is measured for $60$~s using a QuSpin Zero-Field Magnetometer (QZFM) OPM and is sampled at $f_s=200$~Hz using the standard NI LabVIEW QuSpin USB interface.

The shielding efficiencies measured at the shield's center are presented in Table~\ref{tab.shieldstats}. They are $\geq10^6$ in all cases and the $0.2$~Hz evaluations agree to within $10$\% of the expected results from the analytic optimization (in the static limit; see appendix~\ref{app.gaopt}).  As the applied frequency increases, eddy currents induced in the shield increase and tend to enhance $SE_{A/T}$, except for when $SE_T$ reduces between $10$~Hz and $50$~Hz.  This may relate to a resonance in the phase of magnetic fields generated by the oscillating magnetization of the mumetal and induced eddy currents, which are more effective in the transverse shielding case compared to the axial case as the area of the shield perpendicular to the applied field is greater. In addition, in Fig.~\ref{fig.SE}, we show $SE_{A,T}$ measured in $5$~mm increments along the shield's axis measured for $1$~Hz applied fields, sampled over $10$~s at $f_s=10$~kHz at each point using a Mag-13MCZ100 fluxgate magnetometer with a 24-bit Spectramag-6 DAQ (Bartington Instruments). These measurements show good agreement with the OPM data, although the fluxgate noise floor limits the measurement of $SE_{T}$.

A representative residual magnetization profile along the shield's axis inside the MSR is measured using the fluxgate magnetometer and is shown in Fig.~\ref{fig.mag}. The mean absolute field norm between $z=[-L_1/2,L_1/3]$ is $\left(1.2\pm0.1\right)$~nT. The static offset of the fluxgate is accounted for in these measurements by comparing the magnetic field measurements at the center of the shield with the fluxgate reading at three inverted positions, each repeated three times. As the shield has a high $SE_{A,T}$, the residual magnetization will dominate the field profile within the shield when compared to the transmitted field in standard conditions. However, this profile will vary between different recordings since the shield's magnetization is determined by several factors including the background that the shield has experienced, movement of the shield relative to the background, and physical impacts.

\section{Flex-PCB coils}\label{sec.coils}
\begin{table*}
\caption{The benchtop shield contains nine nested flex-PCBs which generate three order $N=1$ uniform harmonics, five $N=2$ linear harmonics, and one $N=3$ quadratic harmonic, with specific variations along the Cartesian unit vectors, $\left(\mathbf{\hat{x}},\mathbf{\hat{y}},\mathbf{\hat{z}}\right)$. The mean field strength, $B_0$, per unit current, $I$, is calculated by averaging the measured field along $\mathbf{\hat{z}}$ between $z=[0,L_1/4]$, except for the $\mathrm{d}B_x/\mathrm{d}x$ and $\mathrm{d}B_y/\mathrm{d}x$ fields which are averaged along $\mathbf{\hat{x}}$ between $x=[0,\rho_1/2]$. Over the same spatial regions, we also evaluate the maximum deviation from the target field, ${\Delta}(B/r^{(N-1)})=B/r^{(N-1)} - B_0$, as a percentage of $B_0$.}
    \centering
\begin{tabular}{| c | c | c || c | c |}
\hline
\multirow{2}{*}{$N$} & \multirow{2}{*}{Target field} & \multirow{2}{*}{Target field harmonic} & Coil efficiency, $B_0/I$ & $\mathrm{max}\left(\left|{\Delta}(B/r^{(N-1)})\right|\right)$ \\
& & & ($\mu$T/(Am$^{(N-1)}$)) & (\%) \\ [2ex]
\hline\hline
\multirow{3}{*}{$1$} & $B_x$ & $B_0\mathbf{\hat{x}}$ & $68.9\pm0.01$ & $0.21\pm0.01$ \\
& $B_y$ & $B_0\mathbf{\hat{y}}$ & $68.7\pm0.01$ & $0.22\pm0.01$ \\
& $B_z$ & $B_0\mathbf{\hat{z}}$ & $74.2\pm0.01$ & $0.44\pm0.01$ \\
\hline
\multirow{5}{*}{$2$} & $\mathrm{d}B_x/\mathrm{d}x$ & $B_0(x\mathbf{\hat{x}}-y\mathbf{\hat{y}})$ & $970\pm10$ & $4\pm2$ \\
& $\mathrm{d}B_y/\mathrm{d}x$ & $B_0(y\mathbf{\hat{x}}+x\mathbf{\hat{y}})$ & $940\pm10$ & $4\pm2$ \\
& $\mathrm{d}B_x/\mathrm{d}z$ & $B_0(z\mathbf{\hat{x}}+x\mathbf{\hat{z}})$ & $430\pm10$ & $2\pm1$ \\
& $\mathrm{d}B_y/\mathrm{d}z$ & $B_0(z\mathbf{\hat{y}}+y\mathbf{\hat{z}})$ & $440\pm10$ & $2\pm1$ \\
& $\mathrm{d}B_z/\mathrm{d}z$ & $B_0(-x\mathbf{\hat{x}}-y\mathbf{\hat{y}}+2z\mathbf{\hat{z}})$ & $1100\pm10$ & $1\pm1$ \\
\hline
$3$ & $\mathrm{d}^2B_z/\mathrm{d}z^2$ & $B_0(-3xz\mathbf{\hat{x}}-3yz\mathbf{\hat{y}}+2z^2\mathbf{\hat{z}})$ & $8800\pm100$ & $4\pm2$ \\
\hline
\end{tabular}
\label{tab.coilperformances}
\end{table*}
\begin{figure}[htb!]
\begin{center}
   \begin{tabular}{c} %% tabular useful for creating an array of images 
         \includegraphics[width=\columnwidth]{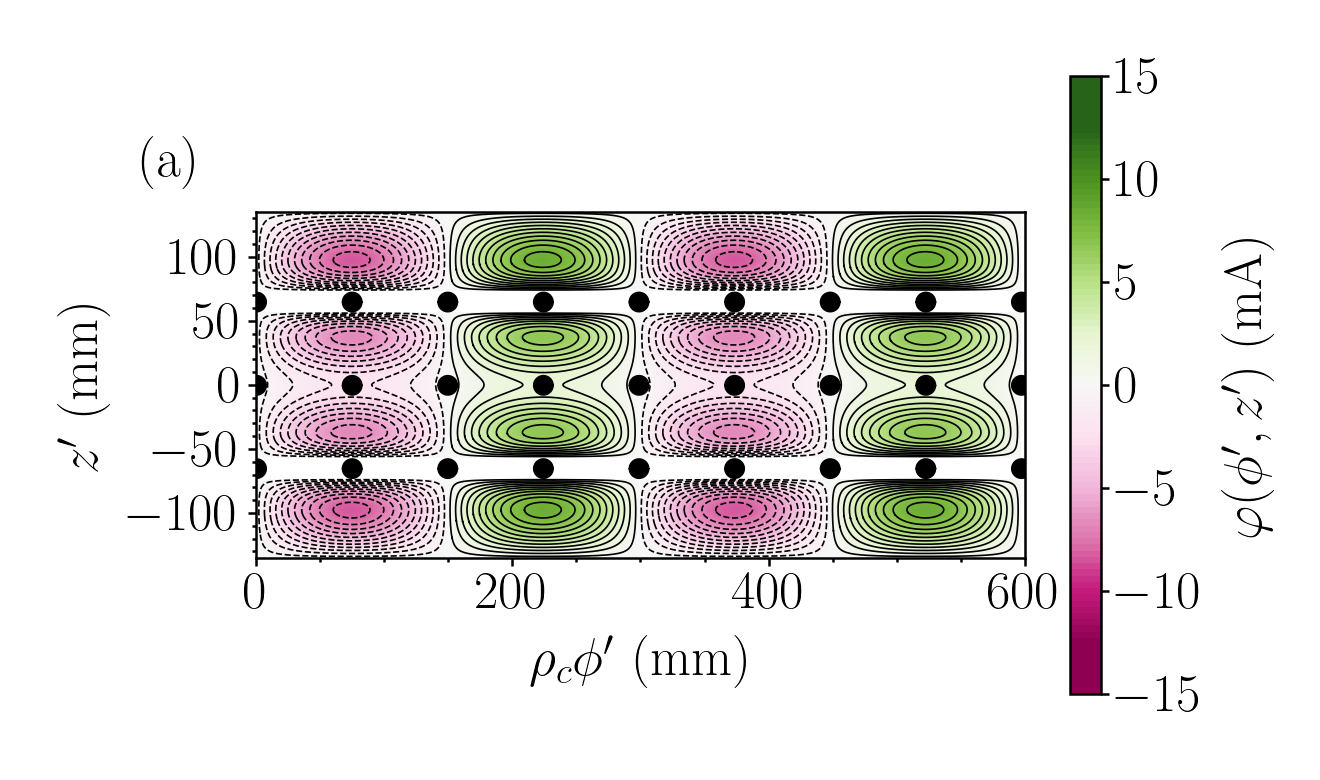} \\
         \vspace{-20pt}
         \includegraphics[width=\columnwidth]{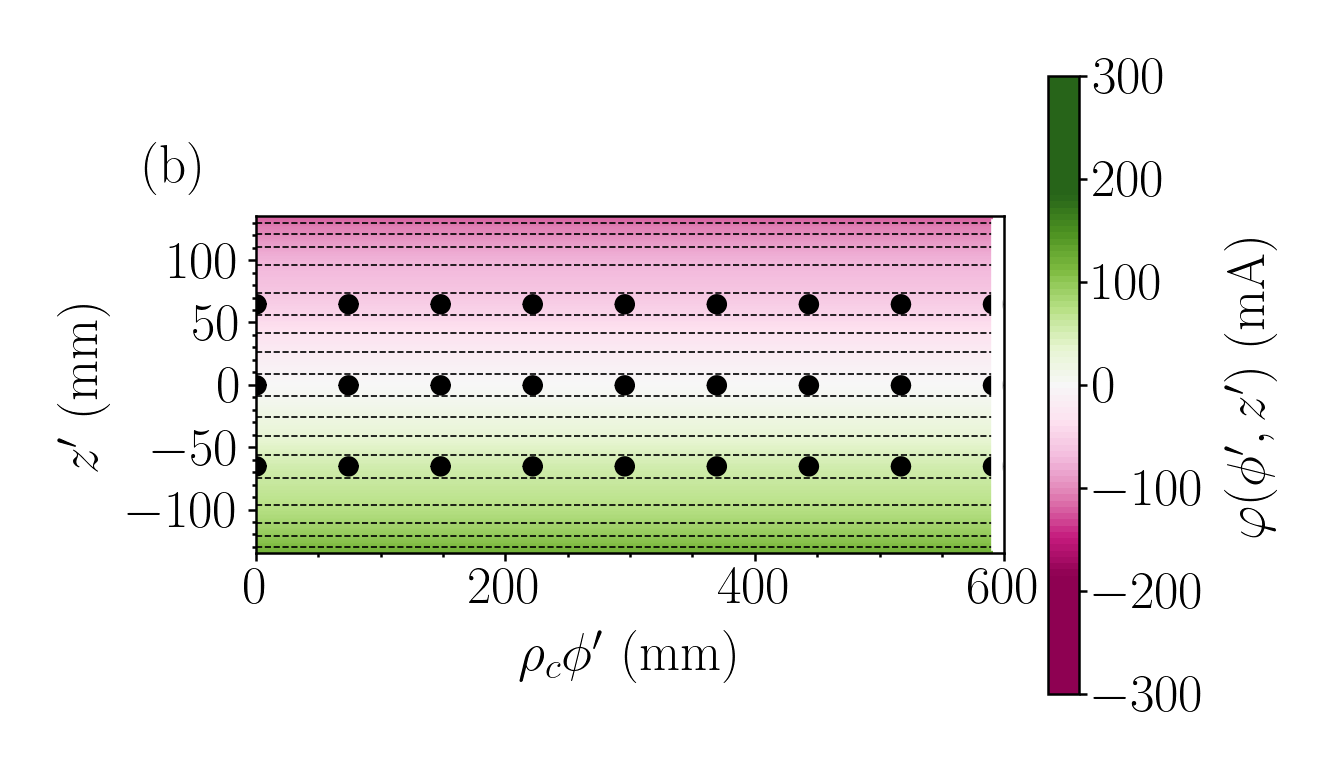}
    \end{tabular}
\end{center}
\vspace{5pt}
\caption{Uniform (a) $\mathrm{d}B_y/\mathrm{d}x$ coil design of radius $\rho_c=94.8$~mm and (b) $B_z$ coil design of radius $\rho_c=92.9$~mm, which extend azimuthally over $\phi'=[0,2\pi]$ and axially over $z'=[-L_c/2,L_c/2]$ where $L_c=270$~mm. Black solid and dashed linestyles show opposite current flow directions; green to white to pink color shows the value of the current flow streamfunction, $\psi\left(\phi',z'\right)$, from positive to zero to negative [scale right]; and black circles show access holes.}
\label{fig.coilpatterns}
\end{figure}
Next, we consider how to null static offsets using active field coils. Firstly, we examine a streamfunction contained on the surface of a coil cylinder of radius $\rho_c$ and length $L_c$, defined by
\begin{align}\label{eq.streamycyl}
    \varphi\left(\phi',z'\right) &= -\sum_{n=1}^{N'}\ \frac{L_c}{n\pi} W_{n0}\cos\left(\frac{n\pi\left(z'-L_c/2\right)}{L_c}\right) + \nonumber \\ &\hspace{-20pt} \sum_{n=1}^{N'}\sum_{m=1}^{M'}\ \frac{L_c}{n\pi} \left(W_{nm}\cos(m\phi')+Q_{nm}\sin(m\phi')\right) \times \nonumber \\ &\hspace{90pt} \sin \left(\frac{n\pi\left(z'-L_c/2\right)}{L_c}\right).
\end{align}
The streamfunction contains modes that are weighted by Fourier coefficients, ($W_{n0}$, $W_{nm}$,  $Q_{nm}$), of order $n\in1,\ldots,N'$ and degree $m\in1,\ldots,M'$.  As the current is confined to the surface of the cylinder, $\nabla\cdot\mathbf{J}\left(z',\phi'\right)=0$, we can express the azimuthal and axial components of the current density in terms of this basis,  $J_{\phi}\left(\phi',z'\right) = \partial\varphi\left(\phi',z'\right)/\partial z'$ and $J_{z}\left(\phi',z'\right) = -\left(1/\rho_c\right)\partial\varphi\left(\phi',z'\right)/\partial\phi'$, respectively~\cite{JacksonEM}.  Coil patterns are generated by using least-squares optimization~\cite{boggs_tolle_1995,floudas1995quadratic,https://doi.org/10.1002/mrm.1910260202} to find optimal values of the Fourier coefficients to generate each target field. The relationship between the magnetic field and the Fourier coefficients is encoded in equations~(37)--(39) in Ref.~\cite{PhysRevApplied.15.054004}.

Here, this method is applied to design nine flex-PCBs to generate nine low-order magnetic field harmonics within the central half length and diameter of the inner shield cylinder. We choose to generate the full set of uniform fields and linear field gradients (see Table~\ref{tab.coilperformances}), and a single quadratic field gradient with respect to axial position, $\mathrm{d}^2B_z/\mathrm{d}z^2$, to help offset the difference between $SE_A$ and $SE_T$. The PCBs are co-centered and co-axial to the shield cylinder and extend over an outer radius $\rho_c=95$~mm and length $L_c=270$~mm.

The streamfunctions and wire patterns which generate the $\mathrm{d}B_y/\mathrm{d}x$ and $B_z$ fields are presented in Fig.~\ref{fig.coilpatterns}. The uniform $\mathrm{d}B_y/\mathrm{d}x$ coil is rolled around its azimuth to form a cylinder of radius $\rho_c=94.8$~mm (Fig.~\ref{fig.shield}b), whereas the uniform $B_z$ coil is rolled into a cylinder of radius $\rho_c=92.9$~mm. The coil patterns have different widths to allow them to be nested inside each other once rolled. The coil patterns are generated by contouring the streamfunction, \eqref{eq.streamycyl}, at evenly spaced levels which span its full domain~\cite{10.1088/0022-3727/34/24/305,10.1088/0022-3727/35/9/303,10.1088/0022-3727/36/2/302}. These patterns are selected according to which best emulates the continuum current~\cite{PhysRevApplied.15.054004} but is manufacturable, i.e. the individual wires are greater than $0.8$~mm apart and do not intersect with the access holes. The PCBs are made of polyimide of $0.26$~mm depth into which copper tracks comprizing the wire patterns are printed and are connected together in series across two flex-PCB layers with vias. The unwanted magnetic fields generated by the connecting tracks are reduced by including tracks on the second PCB layer with opposite current flow. The current pattern which generates the uniform $B_z$ field is composed of current loops in series, which are constructed by soldering bridges across the PCB once it is rolled. The $\mathrm{d}B_y/\mathrm{d}x$ current pattern does not require solder bridges as it does not cross the edge of the PCB. The uniform $B_z$ PCB has a track width of $1.4$~mm to allow $2$~A of current to be passed to produce strong axial biassing (${\sim}150$~$\mu$T, without heating the shield above $40~^{\circ}$C from $20~^{\circ}$C) whereas the remaining PCBs have track widths of $0.4$~mm to allow $500$~mA of current. The flex-PCBs are nested inside a nylon tube and have a radial thickness of $2.5$~mm in situ, including solder bridges.

\begin{figure}[ht!]
\setlength\tabcolsep{5 pt}
\begin{center}
   \begin{tabular}{c c} %% tabular useful for creating an array of images 
         \includegraphics[width=0.45\columnwidth]{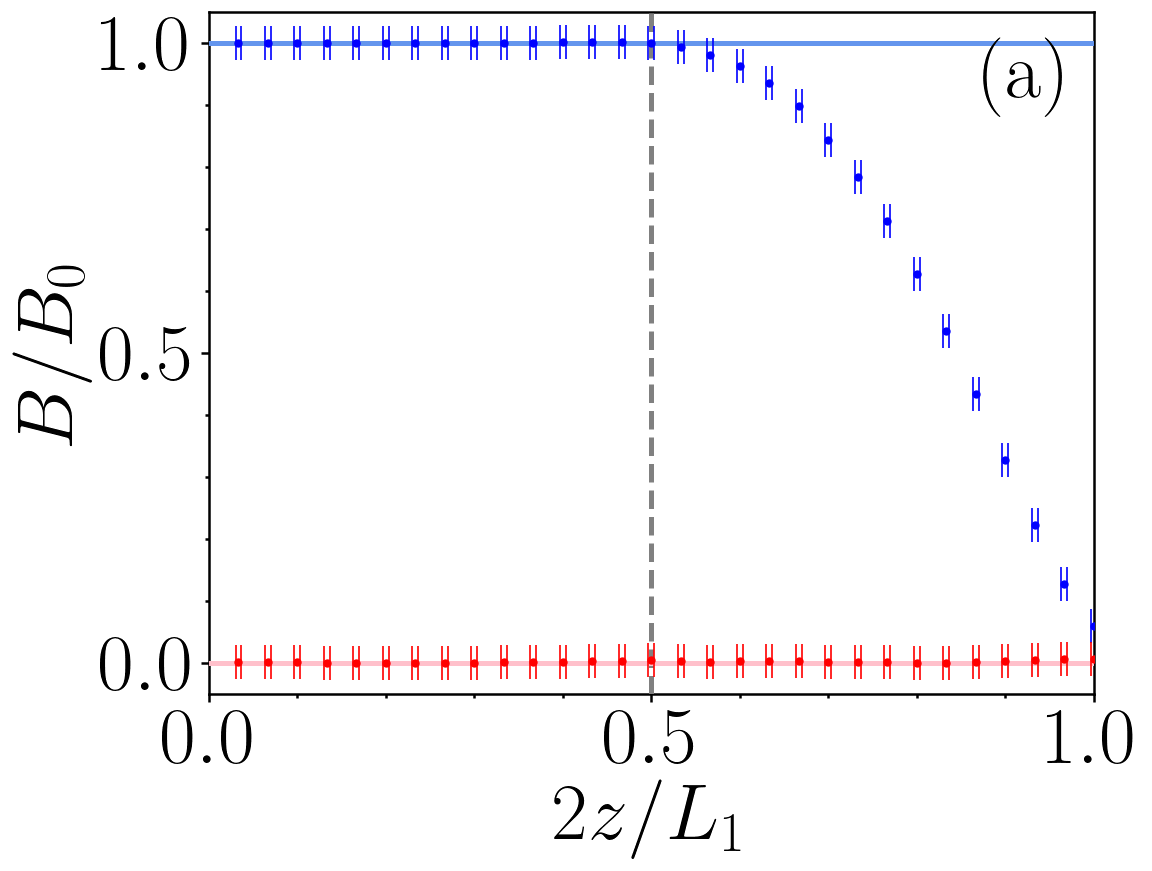} &
         \includegraphics[width=0.45\columnwidth]{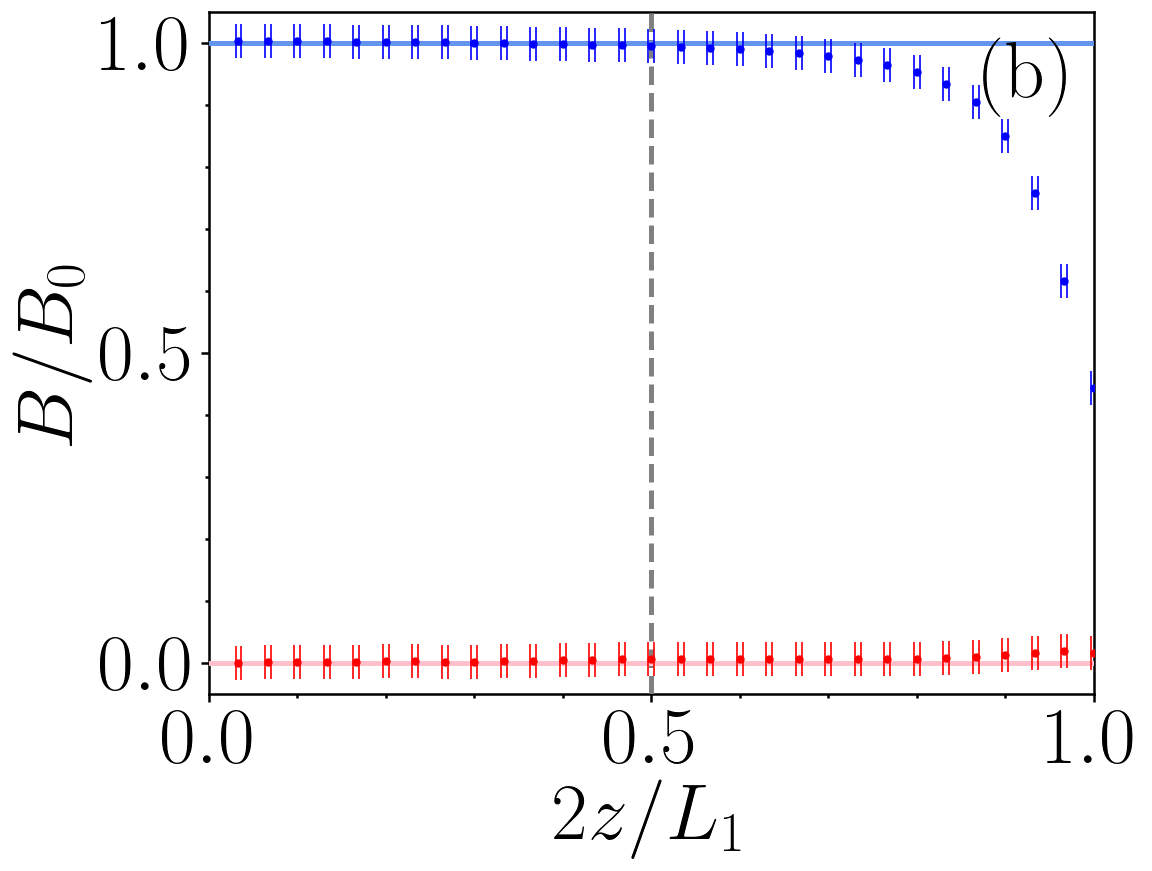} \\
         \includegraphics[width=0.45\columnwidth]{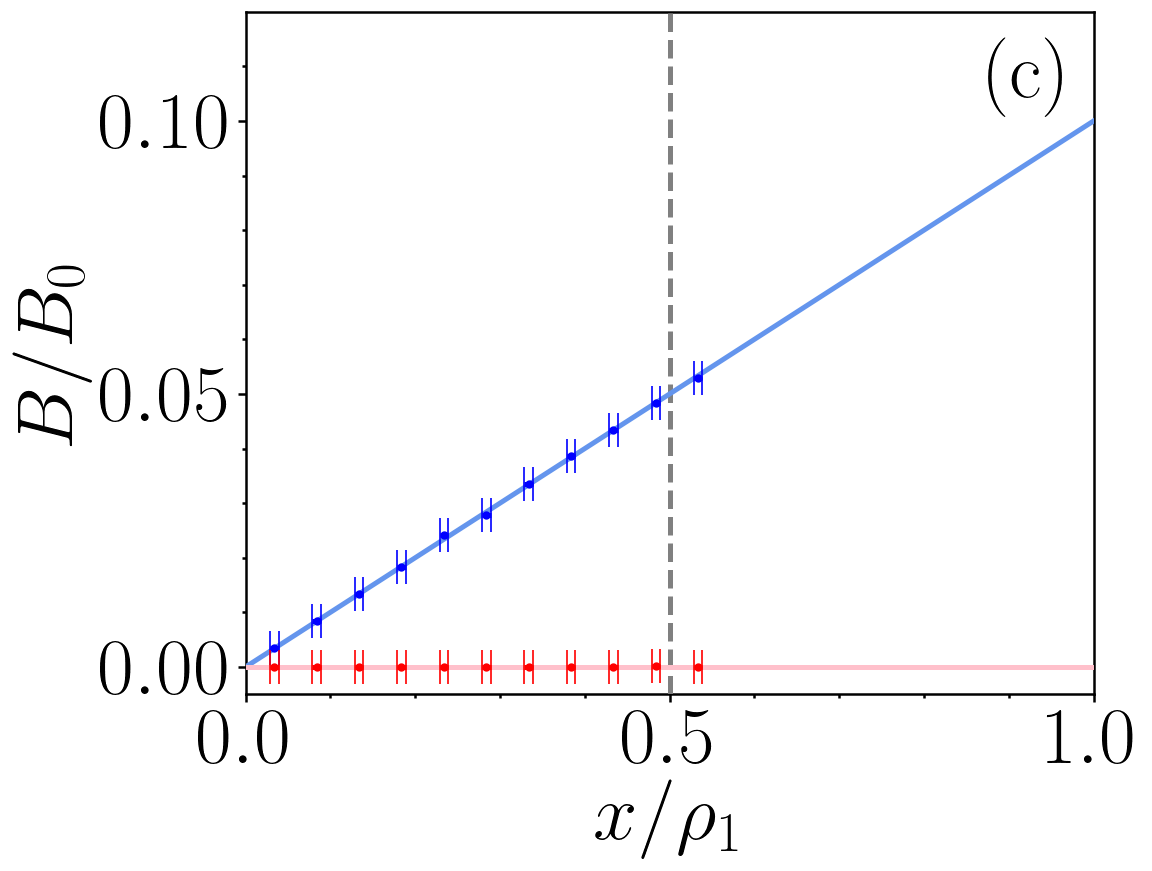} &
         \includegraphics[width=0.45\columnwidth]{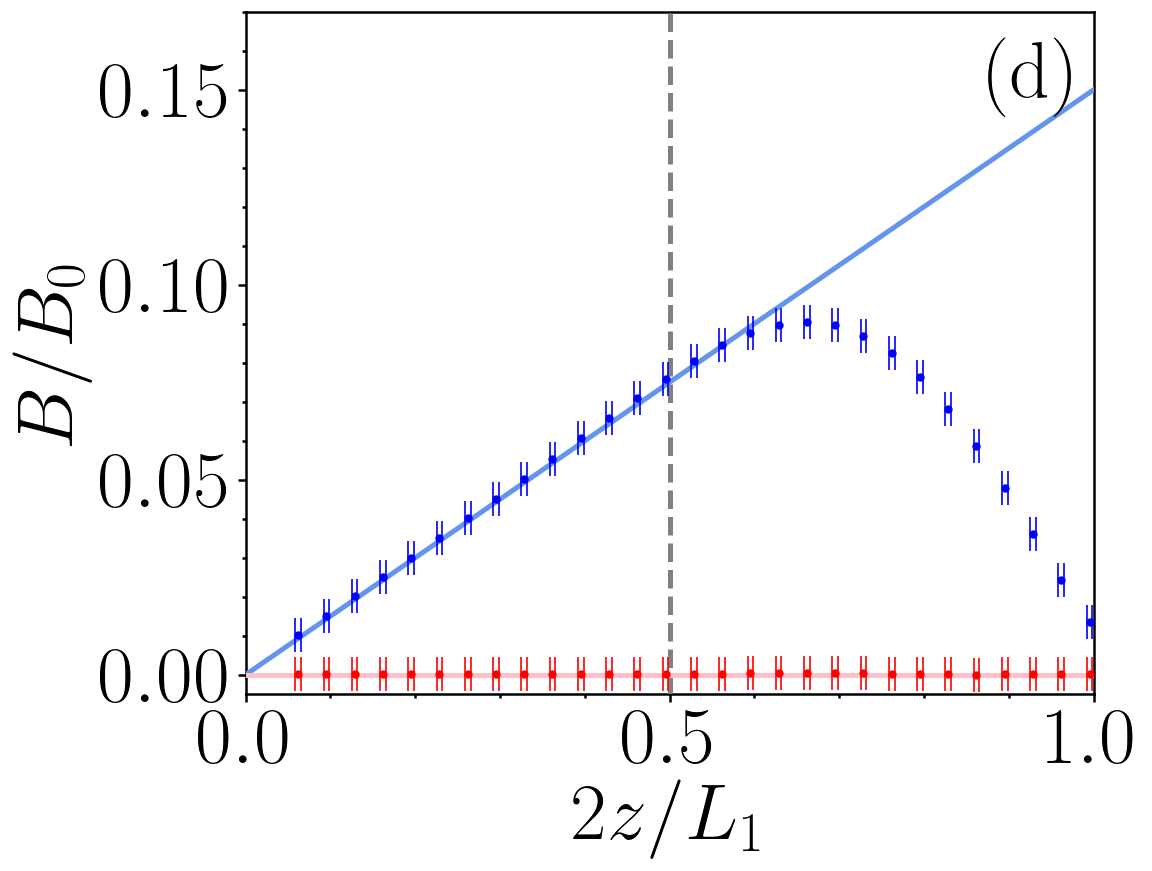} \\
         \includegraphics[width=0.45\columnwidth]{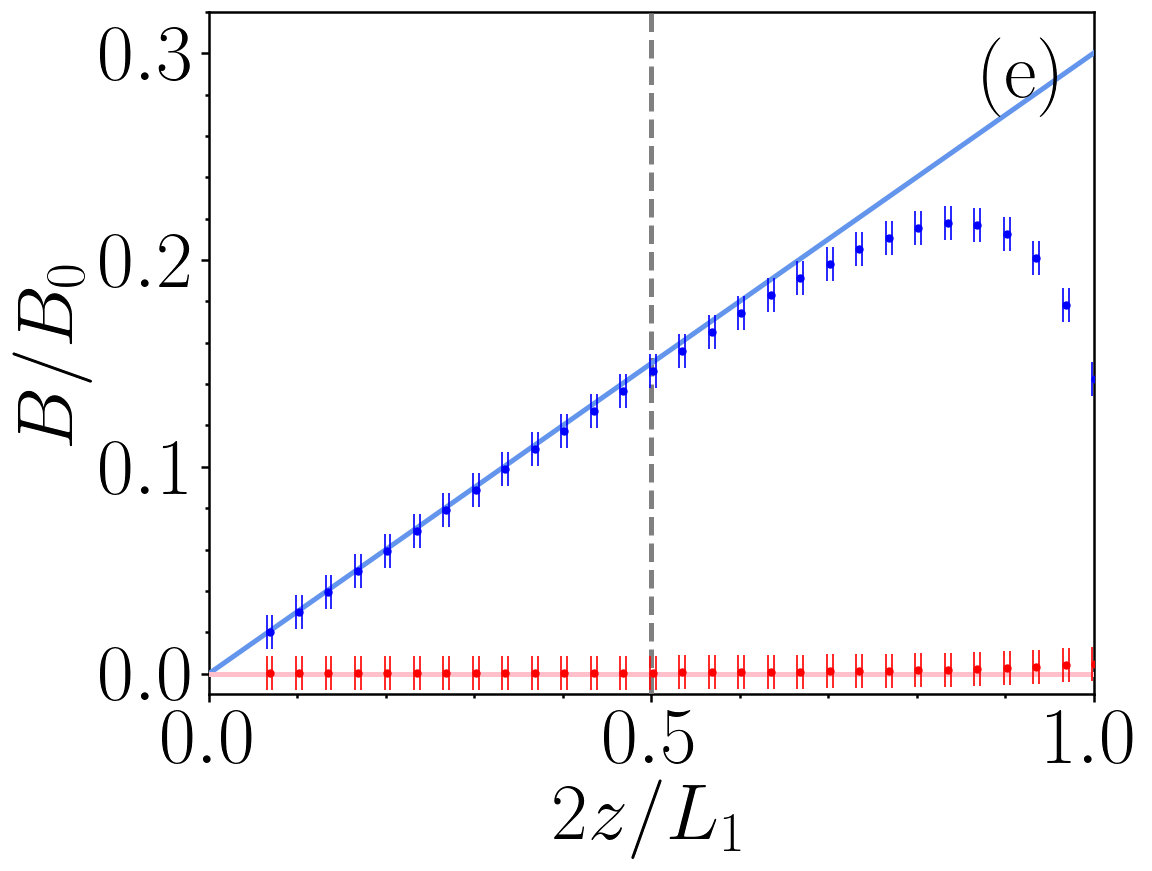} &
         \includegraphics[width=0.45\columnwidth]{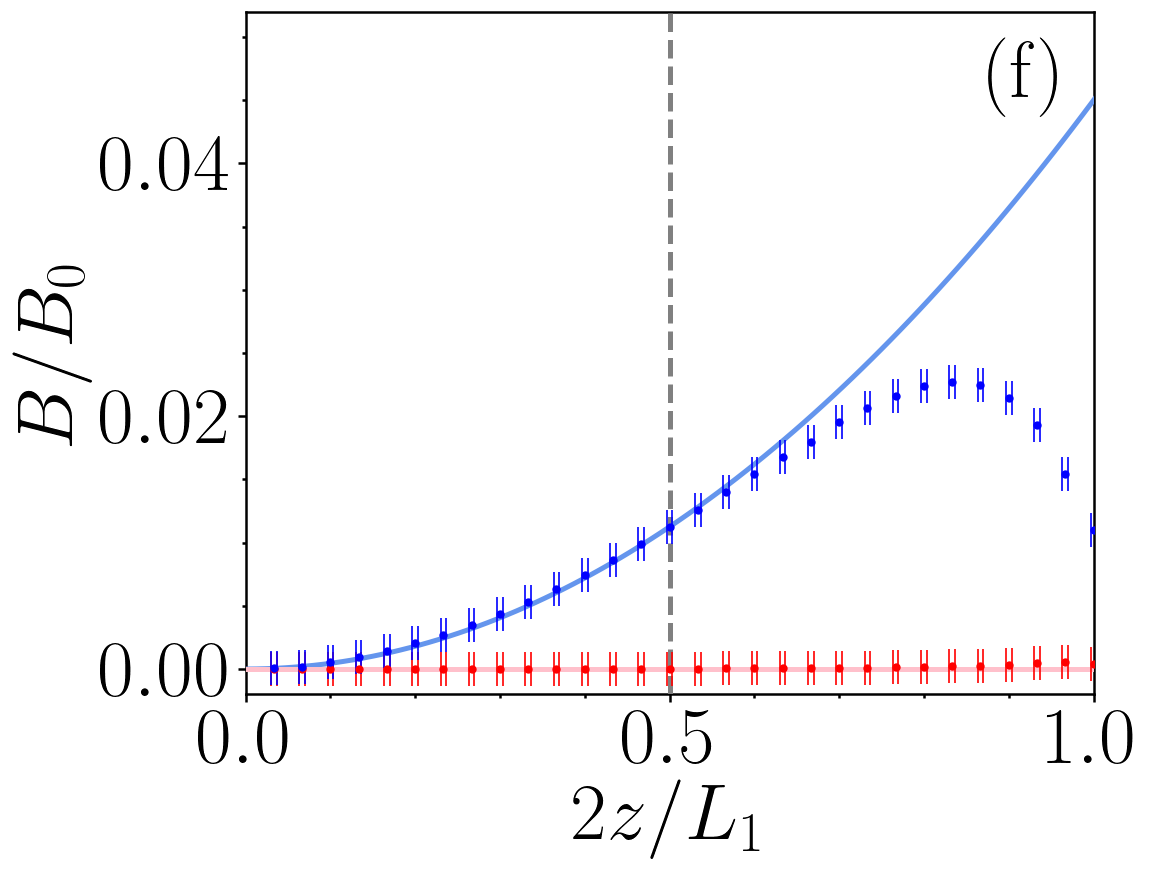}
    \end{tabular}
\end{center}
\caption{Measured magnetic field in the target direction [blue] generated by the uniform (a) $B_y$, (b) $B_z$, (d) $\mathrm{d}B_y/\mathrm{d}z$, (e) $\mathrm{d}B_z/\mathrm{d}z$, and (f) $\mathrm{d}^2B_z/\mathrm{d}z^2$, flex-PCB coils plotted along the $z$-axis of the inner shield cylinder. In (c) the field generated by the $\mathrm{d}B_y/\mathrm{d}x$ coil is plotted along the $x$-axis of the inner shield cylinder. The norm of the magnetic field in the other directions is measured [red] and is expected to be zero. Solid blue lines show perfect representations of the target fields and dashed grey lines show the edge of the target field region.}
\label{fig.coilperformances}
\end{figure}
\begin{figure}[htb!]
\setlength\tabcolsep{5 pt}
\begin{center}
   \begin{tabular}{c c} %% tabular useful for creating an array of images 
         \includegraphics[width=0.45\columnwidth]{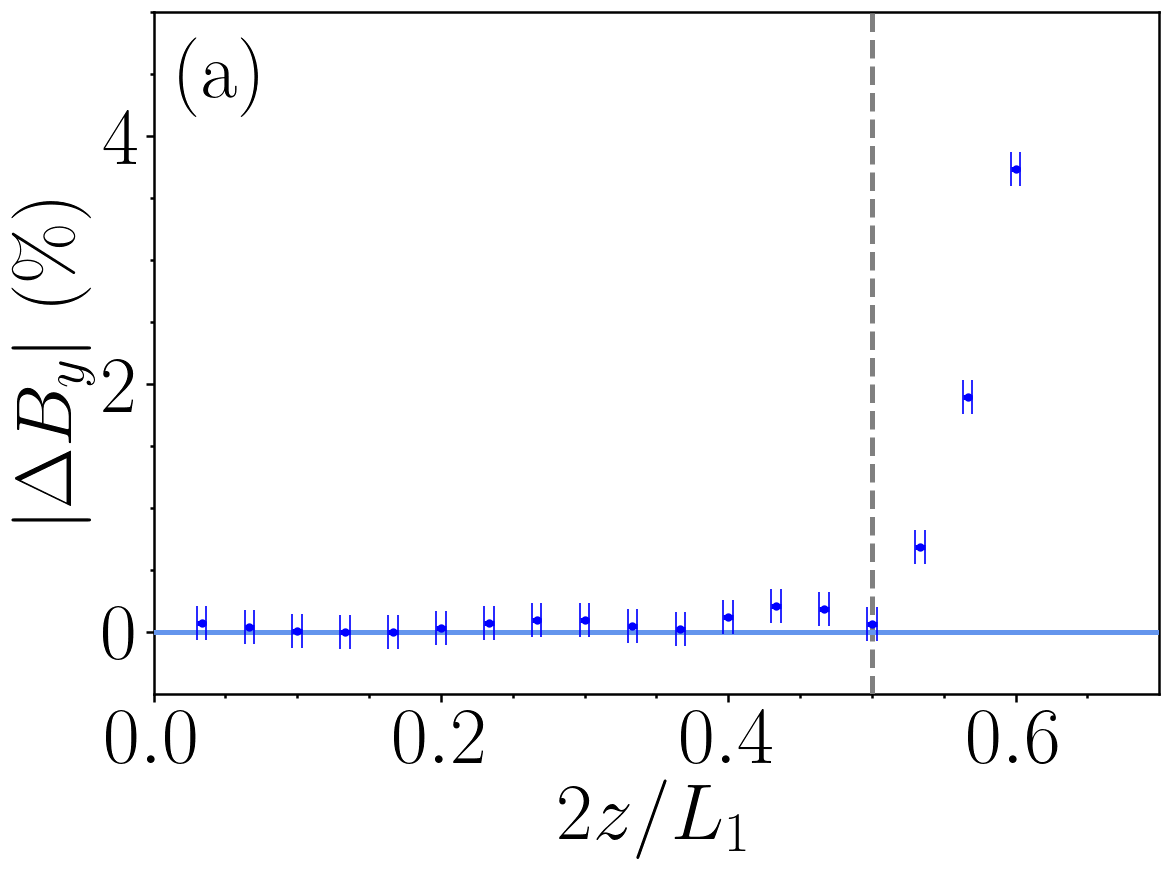} &
         \includegraphics[width=0.45\columnwidth]{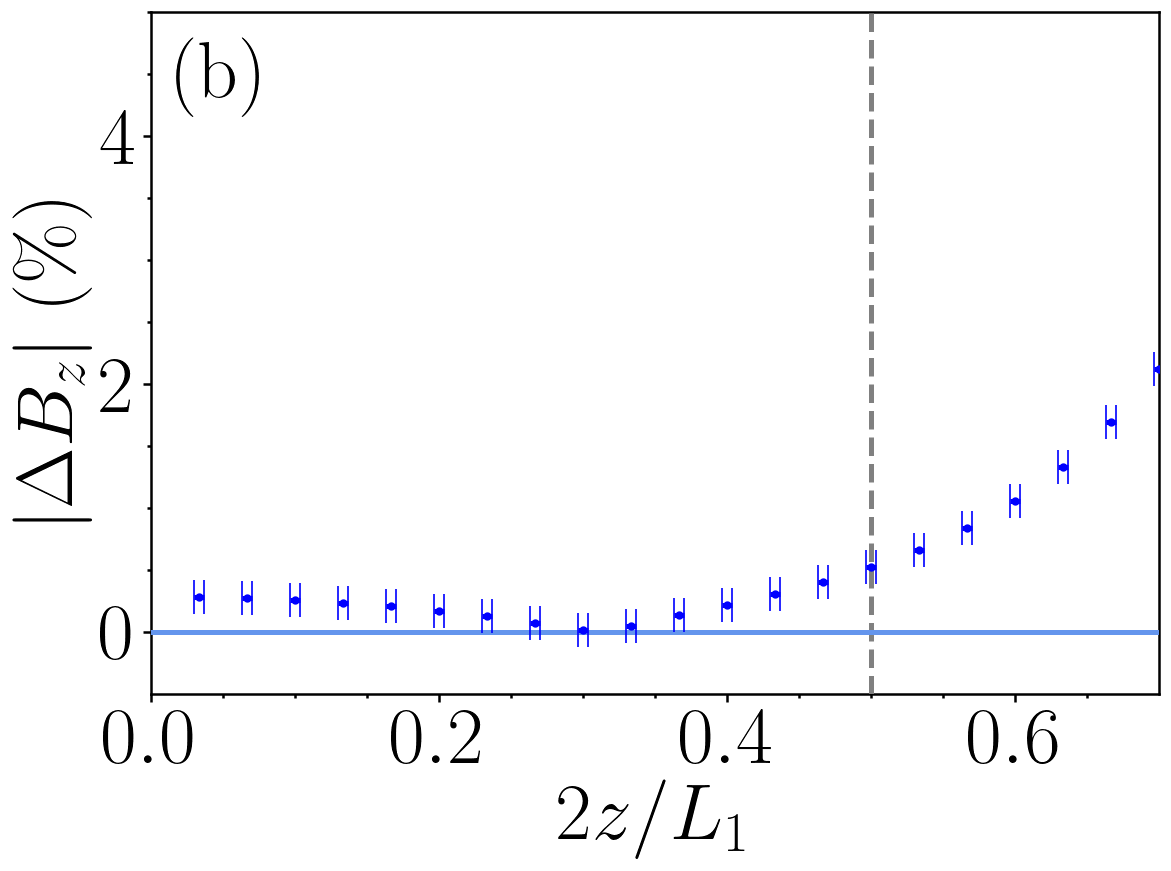} \\
    \end{tabular}
\end{center}
\caption{Deviation between the measured and target fields for the uniform (a) $B_y$ and (b) $B_z$ flex-PCBs plotted versus position along the $z$-axis of the inner shield cylinder. Labelled as Fig.~\ref{fig.coilperformances}.}
\label{fig.coilerrors}
\end{figure}
The magnetic fields generated by each flex-PCB are measured by driving sinusoidal current through each PCB sequentially at a frequency of $1$~Hz for $10$~s and taking the FFT of the measured field. We present the profiles generated by the $B_y$, $B_z$, $\mathrm{d}B_y/\mathrm{d}x$, $\mathrm{d}B_y/\mathrm{d}z$, $\mathrm{d}B_z/\mathrm{d}z$, and $\mathrm{d}^2B_z/\mathrm{d}z^2$ PCBs in Fig.~\ref{fig.coilperformances}, evaluated along the shield's axis, except for the $\mathrm{d}B_y/\mathrm{d}x$ PCB which is evaluated radially as it is designed to generate zero field along the shield's axis. The generated fields show close agreement to the target fields within the target region, and rapidly deviate outside of it, thus minimizing the power consumption required to generate the desired field profile. We examine the deviation from perfect uniformity of the target fields generated by the uniform $B_y$ and $B_z$ PCBs in Fig.~\ref{fig.coilerrors}. The $B_z$ profile deviates more than the $B_y$ profile because of small error fields generated by the connections across the $B_z$ PCB. Notwithstanding this, the fields generated by the uniform field-generating PCBs deviate from target only by $\leq0.5$\% within the target region and compare favorably to other systems optimized in similar contexts~\cite{Hobson_2022}. The remaining PCBs are measured to generate fields which deviate from target by $\leq4$\%; we note that intrinsic deviations are likely to be even smaller as gradient field measurements are highly alignment-sensitive.

\section{Active nulling}\label{sec.hybrid}
\begin{table}[htb!]
\caption{Under applied currents, $I$, the magnitudes of the target fields between $z=[-64,56]$~mm reduce by absolute ratios, $|C|$.}
    \centering
\begin{tabular}{| c || c | c |}
\hline
Target field & Applied current, $I$ & Active nulling ratio, $|C|$ \\
& ($\mu$A) & \\
\hline\hline
$B_x$ & $15.3$ & $11.3$ \\
$B_y$ & $16.9$ & $34.0$ \\
$B_z$ & $-9.4$ & $13.1$ \\
\hline
$\mathrm{d}B_x/\mathrm{d}z$ & $27.4$ & $62.4$ \\
$\mathrm{d}B_y/\mathrm{d}z$ & $ -17.2$ & $6.0$ \\
$\mathrm{d}B_z/\mathrm{d}z$ & $-3.3$ & $9.4$ \\
\hline
$\mathrm{d}^2B_z/\mathrm{d}z^2$ & $-3.4$ & $1.5$ \\
\hline
\end{tabular}
\label{tab.nullperformances}
\end{table}
\begin{figure}[htb!]
\begin{center}
\includegraphics[width=\columnwidth]{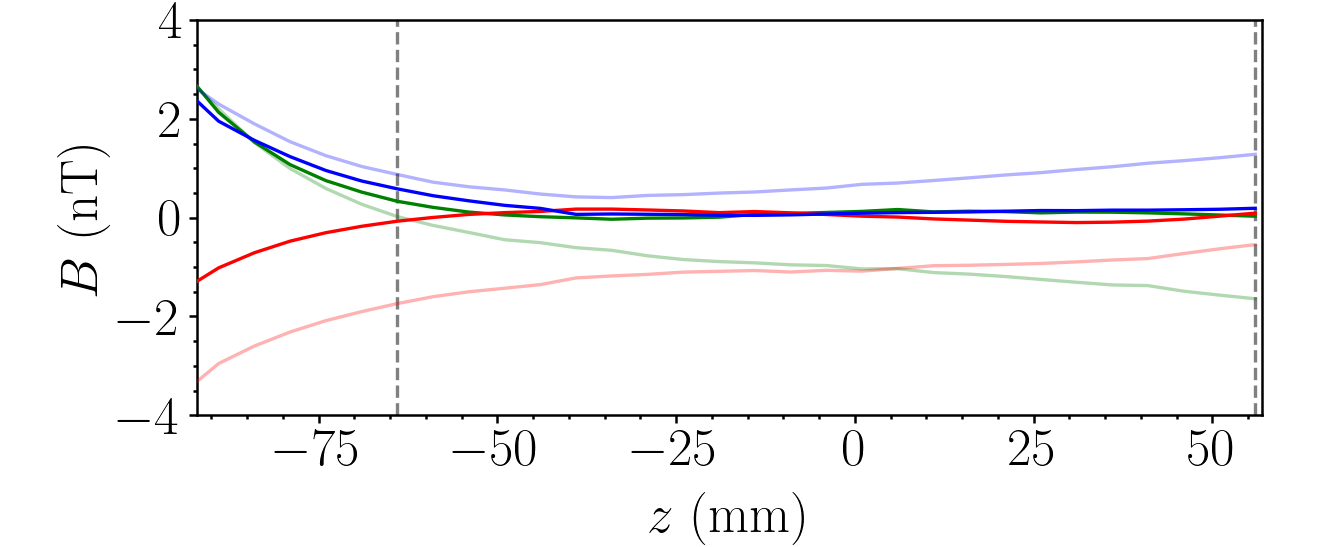}
\end{center}
\caption{$B_x$, $B_y$, and $B_z$ [green, red, and blue] along the $z$-axis of the inner shield cylinder with [solid] and without [light] static active background nulling between $z=[-64,56]$~mm [grey dashed lines].}
\label{fig.null}
\end{figure}
We utilize a QZFM OPM to map the remnant field after degaussing at $5$~mm increments along the shield's axis by calculating $\left(B_x,B_y,B_z\right)$ values required to null the field using the onboard OPM coils using custom MATLAB code which interfaces with the OPM via NI LabVIEW. The same method as described in section~\ref{sec.coils} for the fluxgate is then employed to calculate static offsets. Coil currents are calculated to null the remnant field at $N_\mathrm{null}=25$ points between $z=[-64,56]$~mm, which extends over $40$\% of the inner shield length, by following the methodology outlined in Ref.~\cite{10.1016/j.neuroimage.2021.118401}. Each Cartesian component of the offset-corrected magnetic field is compiled into a list of measurements, $B^\mathrm{mes.}$, of length $\left(3N_\mathrm{null}\right)$. The coil currents required to null these fields may be related to the desired currents in each coil,  $I$, using simple matrix algebra,
\begin{equation}\label{eq.currentcalc}
B^\mathrm{ideal}_{ji}I_{i} = -B^\mathrm{mes.}_j,
\end{equation}
where the matrix $B^\mathrm{ideal}$ contains the expected magnetic field harmonics generated at each sampled coordinate by each flex-PCB used for nulling, assuming unitary current. Here, seven coils are used for nulling since the $\mathrm{d}B_x/\mathrm{d}x$ and $\mathrm{d}B_y/\mathrm{d}x$ coils generate zero field along the $z$-axis, and so this matrix is of dimension $\left(3N_\mathrm{null}{\times}7\right)$. The coil currents are obtained using equation~\eqref{eq.currentcalc} by calculating the pseudo-inverse of $B^\mathrm{ideal}$. The resulting coil currents, which are displayed in Table~\ref{tab.nullperformances}, are at the tens of $\mu$A level due to the low residual field within the shield. A NI-9264 voltage output module is used to generate these currents, which are then amplified using an $8$-channel $\pm10$~V amplifier constructed in-house. The amplifier is experimentally tested to have a noise level $<25$~nV/$\sqrt{\mathrm{Hz}}$ at $5$~Hz. To further ensure that the coil drivers do not add significant noise, each flex-PCB is driven in series with a $47$~k$\Omega$ resistor.

The magnetic field along the $z$-axis pre- and post-null and averaged over two runs is displayed in Fig.~\ref{fig.null}. The active null reduces the mean magnetic field from $1.68$~nT to $0.23$~nT. Re-fitting the field to the harmonic model, we calculate that the norms of the magnitudes of the three target uniform fields and four target gradients are reduced by factors of $19.5$ and $19.8$ after nulling, respectively (see Table~\ref{tab.nullperformances}). The remaining field is dominated by contributions from higher-order field harmonics at the edge of the null region. These harmonics could be alleviated by adding further target field coils to the system to null higher-order gradients or by using retrofitted additional coils, e.g. individually-driven simple building block coils~\cite{DiscretePaper}, for adaptive nulling of residual variations.

\begin{figure}[htb!]
\begin{center}
\includegraphics[width=\columnwidth]{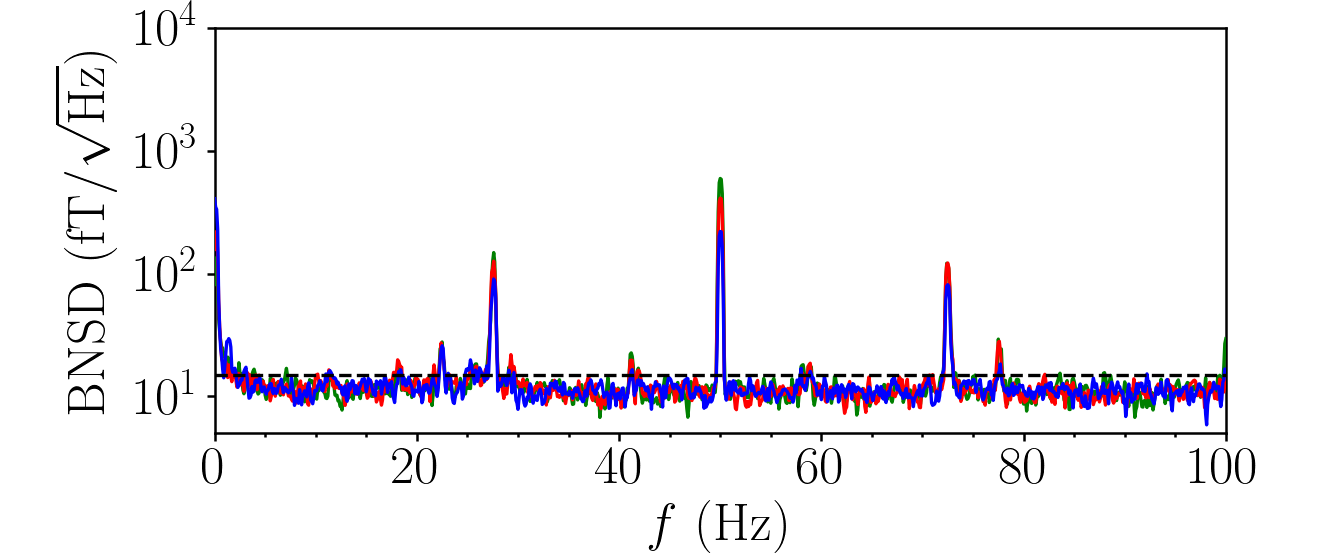}
\end{center}
\caption{Magnetic Noise Spectral Density (BNSD) [noise limit shown by black dashed line] of the $B_x$, $B_y$, and $B_z$ components of the magnetic field [green, red, blue]. Labelled as Fig.~\ref{fig.null}.}
\label{fig.noise}
\end{figure}
The magnetic noise directed along each Cartesian coordinate axis is also calculated at the shield's center by measuring the OPM output over $300$~s while sampling at $f_s=1.2$~kHz. This is displayed in Fig.~\ref{fig.noise}. Generally, the noise is limited by the OPM noise floor, $15$~fT/$\sqrt{\mathrm{Hz}}$, and so the shield-induced Johnson noise is less than this value. The noise floor peaks at $\sim750$~fT/$\sqrt{\mathrm{Hz}}$ at $f=50$~Hz due to mains electrical noise, with projections at $f=[28,72]$~Hz due to the OPM powerline.

\section{Conclusion}

In this article, we have developed a magnetic field control system by utilizing optimized arrangements of mumetal magnetic shielding in conjunction with a set of nine internal flex-PCBs. Experimental measurements show that the flex-PCBs generate target magnetic fields with $<4\%$ error along measured target regions which extend over $50$\% of either the inner shield diameter or length. The flex-PCBs are demonstrated to enhance the effectiveness of the passive shields. They reduce the mean static magnetic field norm to $0.23$~nT over $40$\% of the inner shield's axis in a typical laboratory environment, which is a $7.3$-fold improvement compared to the performance of the passive shields alone. Our results also demonstrate that the active coil nulling may be used without introducing significant magnetic noise. The magnitude of magnetic noise measured by a commercially available zero-field OPM was evaluated to be broadly at the sensor's $15$~fT/$\sqrt{\mathrm{Hz}}$ noise limit over much of the spectrum from $0$ to $100$~Hz.

The design, manufacturing, and experimental processes laid out in this work will be instructive to the development of magnetic shields in several application spaces outside of benchtop shielding. Larger shields designed using the same methods could be used similarly to lightweight MSRs~\cite{10.1038/s41598-022-17346-1} for recording muscle~\cite{BROSER2021102490} or gut activity~\cite{10.1016/j.compbiomed.2020.104169}. Since these shields would require larger access holes, additional field-generating systems would be required to reduce leakage fields. Such coil systems may also supplement existing shielding, enabling its partial removal, e.g. for weight reduction in spacecraft~\cite{6038424}.

\section*{Acknowledgements}
We acknowledge the support of the UK Quantum Technology Hub Sensors and Timing (EP/T001046/1) and from Innovate UK Project 44430 MAG-V: Enabling Volume Quantum Magnetometer Applications through Component Optimization \& System Miniaturization.

\section*{Declarations}
P.J.H, M.P, D.S, M.B, R.B, and M.F have a worldwide patent (WO/2021/053356) which includes the coil design technique applied in this work. B.S, P.P, J.C, and D.W are employees of Magnetic Shields Limited (MSL), who sell benchtop shields commercially. S.J, D.H, and R.H are ex-employees of MSL. N.H, D.W, M.B, and R.B hold founding equity in Cerca Magnetics Limited, who commercialize OPM technology. C.M, A.D, and T.S declare no competing interests. \\

MSL have made the benchtop shield available for purchase and may be contacted via: \href{mailto:enq@magneticshields.co.uk}{\underline{enq@magneticshields.co.uk}}. \\

This work has been submitted to the IEEE for possible publication. Copyright may be transferred without notice, after which this version may no longer be accessible. \\

All supporting data may be made available on request.

\bibliographystyle{IEEEtran}
\bibliography{literatur}

\begin{appendices}

\section{Passive shields optimization}\label{app.gaopt}

Sumner~\cite{Sumner_1987} generated simple approximations for the static shielding effectiveness, $SE$, of $N_s$ nested layers of cylindrical passive magnetic shielding of a fixed high relative magnetic permeability, $\mu_r$, and length greater than radius, $L_i>\rho_i$, in the flux-shunting dominated limit~\cite{doi:10.1002/9780470268483.app2}. These are
\begin{align}\label{eq.SE}
SE &= 1 + \sum^{N_s}_{i=1}SE_i + \sum^{N_s-1}_{i=1}\sum^{N_s}_{j>i}SE_{i}SE_{j}F_{ij} + \nonumber \\ &\hspace{20pt} \sum^{N_s-2}_{i=1}\sum^{N_s-1}_{j>i}\sum^{N_s}_{k>j}SE_{i}SE_{j}SE_{k}F_{ij}F_{jk} + \ldots + \nonumber \\ &\hspace{100pt} SE_{N_s}\prod^{N_s-1}_{i=1}SE_{i}F_{i(i+1)},
\end{align} 
where $SE_{i}$ and $F_{ij}$ are replaced by different functions depending on whether the shielding efficiency axial, $SE_A$, or transverse, $SE_T$, to the shield's axis is to be calculated. In the transverse case, they are replaced by
\begin{align}\label{eq.SE_T}
SE_{Ti} &= \frac{{\mu_r}{d_i}}{2\rho_i}, \\ T_{ij} &= 1 - \left(\frac{\rho_i}{\rho_j}\right)^2.
\end{align} 
Whereas, in the axial case, they are replaced by
\begin{align}\label{eq.SE_A}
SE_{Ai} &= 1 + \frac{{\mu_r}{d_i}}{2\rho_i}\left(\frac{K_i}{1 + \rho_i + 0.85\rho_i^2/3}\right), \\ A_{ij} &= 1 - \left(\frac{L_i}{L_j}\right),
\end{align}
where,
\begin{align}\label{eq.K}
K_i &= \left(1.7 - \frac{1}{\rho_i} - 1.35\left( 1 + \frac{1}{4\rho_i^3}\right)\right) \times \nonumber \\ &\hspace{10pt} \left(\ln{\left(\rho_i + \sqrt{1 + \rho_i^2}\right)} - 2\left(\sqrt{1 + \frac{1}{\rho_i^2}} - \frac{1}{\rho_i}\right)\right).
\end{align} 
Here, we will use equations~\eqref{eq.SE_T}~and~\eqref{eq.SE_A} with $\mu_r=40000$ as objective functions for a genetic algorithm. An additional objective function is the total volume of shielding material in the cylindrical shield wall and circular end caps of the shielding layers,
\begin{equation}\label{eq.Vol}
V = \sum^{N_s}_{i=1} 2\pi d_i \rho_i^2 + \pi \left(2d_i\rho_i-d_i^2\right) \left(L_i - 2d_i\right).
\end{equation}

In the optimization system, as detailed in the main text, we fix the geometry of the inner shield cylinder and the thicknesses of all the shielding layers. The fixed values imposed in the optimization are therefore
\begin{equation}\label{eq.gaobj}
    \texttt{fixed~vals.}
    \begin{cases}
    \rho_1 = 100~\mathrm{mm}, \\ 
    L_1 = 300~\mathrm{mm}, \\ 
    d_1 = 0.5~\mathrm{mm}, \\
    d_i = 1.5~\mathrm{mm} \mathrm{~for~}i\in[2,3,4].
    \end{cases}
\end{equation}
To ensure that the shielding layers have sufficient clearance for manufacture and so the shield fits on a laboratory benchtop, the boundaries of the search domain are constrained to
\begin{equation}\label{eq.searchdomain}
    \texttt{search~bounds.}
    \begin{cases}
    \rho_{i+1} - \rho_{i} > 5~\mathrm{mm}\mathrm{~for~}i\in[1,2,3], \\
    L_{i+1} - L_{i} > 5~\mathrm{mm}\mathrm{~for~}i\in[1,2,3], \\
    \rho_4 \leq 150~\mathrm{mm}, \\
    L_4 \leq 600~\mathrm{mm},
    \end{cases}
\end{equation}
The objectives of the optimization procedure are therefore
\begin{equation}\label{eq.objective}
    \texttt{obj.}
    \begin{cases}
    \max SE_T\left(\rho_2,\rho_3,\rho_4,L_2,L_3,L_4\right), \\
    \max SE_A\left(\rho_2,\rho_3,\rho_4,L_2,L_3,L_4\right), \\
    \min V\left(\rho_2,\rho_3,\rho_4,L_2,L_3,L_4\right).
    \end{cases}
\end{equation}
\begin{figure}[htb!]
\centering
\begin{center}
\includegraphics[width=0.9\columnwidth]{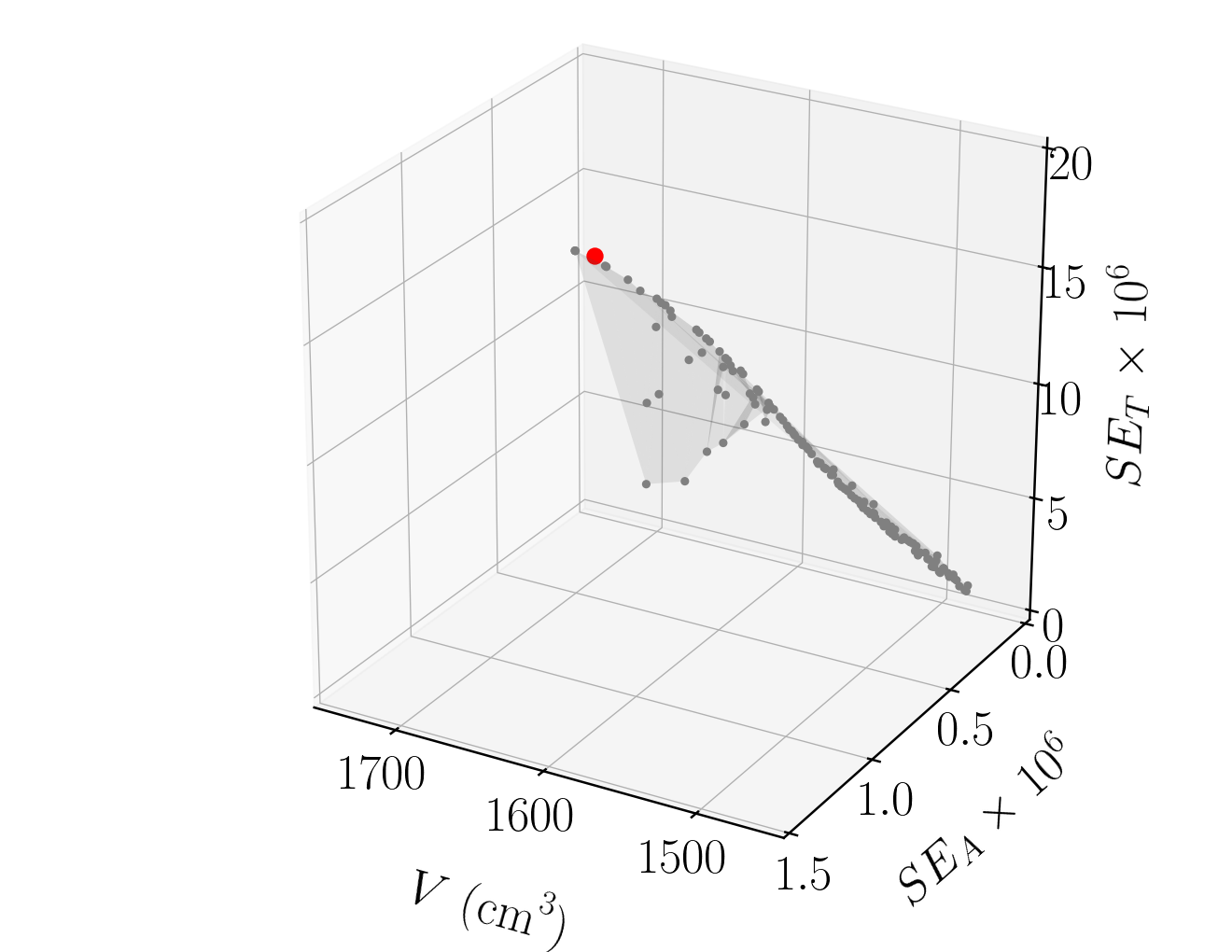}
\end{center}
\caption{Pareto front [grey scatter] where the shielding efficiencies axial, $SE_A$, and transverse, $SE_T$, to the shield's axis are maximized and total volume of shielding material, $V$, is minimized. Pareto-optimal solutions are filtered to maximize $SE_A$ [chosen solution red].}
\label{fig.geneticalgorithm}
\end{figure}
The NSGA-II genetic algorithm is utilized to find solutions for the objective functions laid out in equation~\eqref{eq.objective}, subject to the search domain constraints outlined in equation~\eqref{eq.searchdomain}. The NGSA-II control parameters used match those in Ref.~\cite{DiscretePaper}, except that the population size set to $250$. Convergence is achieved after $112$ generations of the algorithm, requiring $28000$ evaluations of each objective function. This takes $42.6$ seconds on a MacBook Pro16,1 which is equipped with a 6-Core Intel Core i7 2.6 GHz processor and 16 GB of DDR4 RAM. The output Pareto front is presented in Fig.~\ref{fig.geneticalgorithm}. The solution with the highest axial shielding efficiency is selected, where the axial and transverse shielding efficiencies are $SE_A=1.1 \times 10^6$ and $SE_T=19 \times 10^7$, respectively, for fixed $\mu_r=40000$. These values deviate from the lowest frequency ($0.2$ Hz) experimentally measured values in Table~\ref{tab.shieldstats} by $10$\%, although are highly sensitive to the value of initial relative permeability selected in the optimization. Thus, discrepancies are anticipated as the analytic formulae are approximations in the static limit and assume fixed permeability of each layer. Conversely, in the experimental measurements, the inner layers of mumetal will experience a reduced field magnitude, changing their effective permeability~\cite{ANDALIB2017139}.
\end{appendices}
\end{document}